\documentclass[aps,prl,superscriptaddress,twocolumn,groupedaddress,showpacs]{revtex4}
\usepackage[utf8]{inputenc}

\usepackage{amsmath}

\usepackage{hyperref}
\usepackage{amssymb}
\usepackage{graphicx,color}

\usepackage{amsmath,amsthm}

\usepackage{mathrsfs}

\usepackage{bm}

\newcommand{\ccirc}{\kern0.2ex\vcenter{\hbox{$\scriptstyle\circ$}}\kern0.2ex}

\def\be{\begin{eqnarray}}
\def\ee{\end{eqnarray}}

\newcommand{\ch}{\mathcal H}

\begin{document}


\title{Efficient Simulation of Loop Quantum Gravity \---- \\ A Scalable Linear-Optical Approach}

\author{Lior Cohen}
\affiliation{%
Hearne Institute for Theoretical Physics, and Department of Physics and Astronomy, Louisiana State University, Baton Rouge, Louisiana 70803, USA.
}

\author{Anthony J. Brady}
\affiliation{%
Hearne Institute for Theoretical Physics, and Department of Physics and Astronomy, Louisiana State University, Baton Rouge, Louisiana 70803, USA.
}

\author{Zichang Huang}
\affiliation{Department of Physics, Center for Field Theory and Particle Physics, and Institute for Nano-
electronic devices and Quantum computing, Fudan University, Shanghai 200433, China}
\affiliation{State Key Laboratory of Surface Physics, Fudan University, Shanghai 200433, China}

\author{Hongguang Liu}
\affiliation{Center for Quantum Computing, Pengcheng Laboratory, Shenzhen 518066, China}

\author{Dongxue Qu}
\affiliation{Department of Physics, Florida Atlantic University, 777 Glades Road, Boca Raton, FL 33431, USA}

\author{Jonathan P. Dowling}\thanks{Deceased}
\affiliation{%
Hearne Institute for Theoretical Physics, and Department of Physics and Astronomy, Louisiana State University, Baton Rouge, Louisiana 70803, USA.
}
\affiliation{%
	NYU-ECNU Institute of Physics at NYU Shanghai, 3663 Zhongshan Road North, Shanghai, 200062, China.
	}
\affiliation{%
CAS-Alibaba Quantum Computing Laboratory, CAS Center for Excellence in Quantum Information and Quantum Physics,     University of Science and Technology of China, Shanghai 201315, China.
    }
\affiliation{%
National Institute of Information and Communications Technology,
4-2-1, Nukui-Kitamachi, Koganei, Tokyo 184-8795, Japan
    }

\author{Muxin Han}\thanks{Corresponding author:  hanm@fau.edu}
\affiliation{Department of Physics, Florida Atlantic University, 777 Glades Road, Boca Raton, FL 33431, USA}
\affiliation{Institut f\"ur Quantengravitation, Universit\"at Erlangen-N\"urnberg, Staudtstr. 7/B2, 91058 Erlangen, Germany}


\begin{abstract}
The problem of simulating complex quantum processes on classical computers gave rise to the field of quantum simulations. Quantum simulators solve problems, such as Boson sampling, where classical counterparts fail.   
In another field of physics, the unification of general relativity and quantum theory is one of the greatest challenges of our time. One leading approach is Loop Quantum Gravity (LQG).
Here, we connect these two fields and design a linear-optical simulator such that the evolution of the optical quantum gates simulates the spinfoam amplitudes of LQG. It has been shown that computing transition amplitudes in simple quantum field theories falls into the class BQP -- which strongly suggests that computing transition amplitudes of LQG are classically intractable. Therefore, these amplitudes are efficiently computable with universal quantum computers which are, alas, possibly decades away. We propose here an alternative special-purpose linear-optical quantum computer, which can be implemented using current technologies. This machine is capable of efficiently computing these quantities.   
This work opens a new way to relate quantum gravity to quantum information and will expand our understanding of the theory. 

\end{abstract}


\maketitle











 




\textit{Introduction.}\----Linear optics promises a great opportunity to implement and execute quantum protocols in order to accomplish quantum computational and quantum information processing tasks \cite{knill2001scheme,kok2007linear}. Photons are also the fastest qubits -- a crucial property for quantum communication \cite{bennet1984quantum} -- and their easy manipulation makes them ideal for quantum sensing applications as well \cite{motes2015linear,vanmeter2007general}.
In addition, linear optics has been shown to be useful for entangled-state preparation \cite{wang2018multidimensional} and quantum circuit preparation \cite{reck1994experimental}, with applications to, e.g., one-way quantum computing \cite{raussendorf2001one,walther2005experimental,istrati2020sequential}. 

Simulations of complex quantum systems are inefficient while running on conventional computers \cite{feynman1999simulating}. 
However, efficient quantum simulations are within reach, if run on near-term quantum computers or quantum simulators \cite{feynman1985quantum,lloyd1996universal}. Several protocols for efficient quantum simulations are realizable using linear optics \cite{cerf1998optical,howell2000reducing,lu2009demonstrating}. One distinguished example \---- Boson sampling \---- has been demonstrated \cite{tillmann2013experimental,spring2013boson,wang2017high,zhong201812}. That work has led to a Boson-sampling-inspired algorithm for simulating vibrational states of molecules \cite{huh2015boson}. 

Quantum simulators also have applications in fundamental physics such as, e.g., efficient simulation of quantum field theories \cite{jordan2012algorithm,jordan2018bqp,preskill2018qftQuComp}. Indeed,  S. P. Jordan et al. \cite{jordan2018bqp} has recently shown that computing even simple quantum field-theoretic transition amplitudes falls within the computational complexity class BQP. This result strongly suggests that computing the transition amplitudes of, say, Loop Quantum Gravity (LQG) \---- a more complicated quantum field theory \---- also falls into this class. 

We therefore anticipate that LQG amplitudes, specifically spinfoam amplitudes, are efficiently calculable on universal quantum computers -- which may be decades away. Contrariwise, in this work, we design a special-purpose linear-optical quantum computer able to compute the spinfoam amplitudes of LQG efficiently. Since the spinfoam amplitudes are related to many key issues in LQG \---- such as the semiclassical limit, the continuum limit, and many key physical predictions -- our results may shed light on fundamental aspects of quantum gravity. 


LQG is a background-independent and non-perturbative approach to the theory of quantum gravity \cite{thiemann2008modern,han2007fundamental,ashtekar2004background}. As LQG analogs of Feynman path integrals for quantum gravity, spinfoam amplitudes are transition amplitudes for the evolution of LQG quantum geometry states \cite{reisenberger1997sum,rovelli2014covariant,perez2013spin}. The spinfoam amplitude plays the central role in the covariant dynamics of LQG in 3+1 dimensions. 



\begin{figure}[t]
  \begin{center}
    \includegraphics[width = 0.35\textwidth]{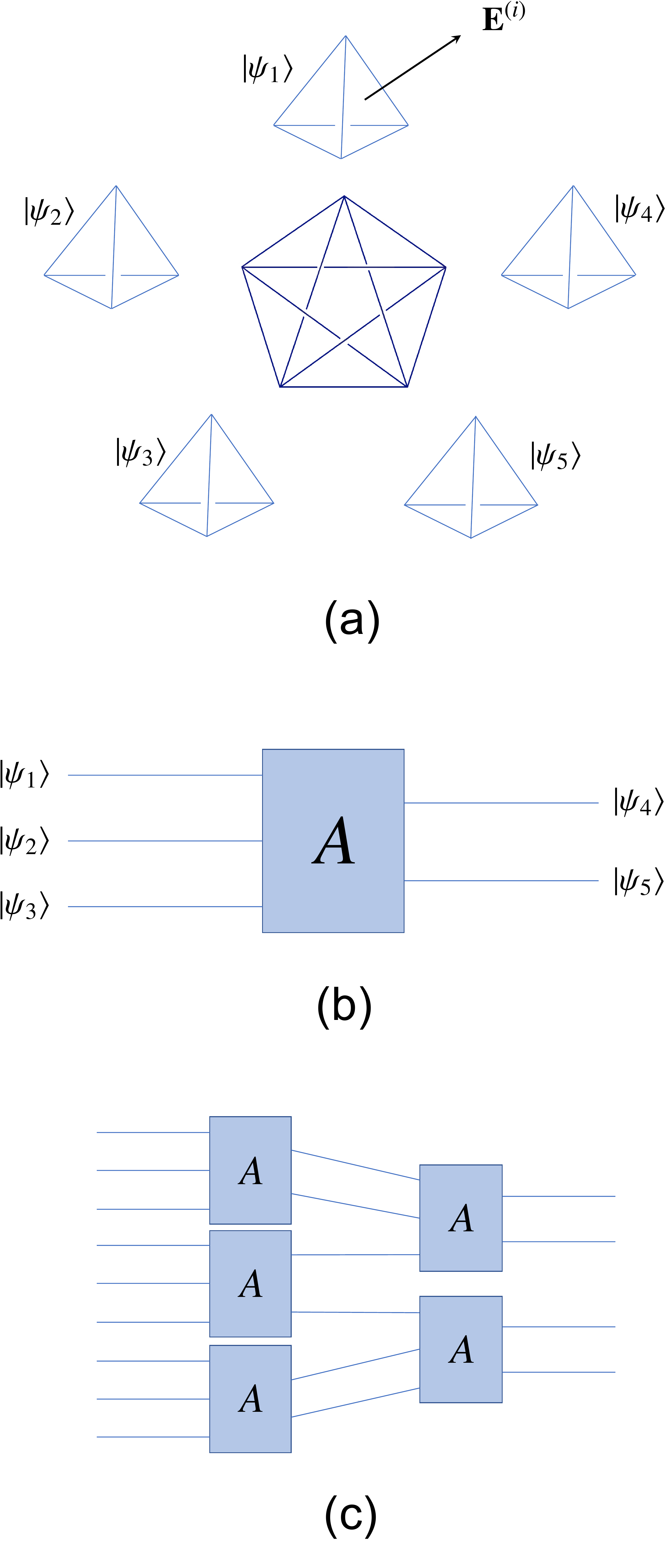}
  \end{center}
  \caption{(a) A four-simplex whose boundary is made by five tetrahedra. Each tetrahedron is quantized to $|\psi_i\rangle\in{\mathcal H}_{\rm tet}$ ($i=1,\cdots,5$). ${\bf E}^{(i)}$ is the oriented area vector of the $i^{\rm th}$ tetrahedron face, and is quantized as quantum angular momenta. (b) The quantum gate $A$ with three input quantum tetrahedra and two output. We note that at least three qubits are needed to operate this non-unitary gate and even more if a unitary expansion of the gate is used. The spinfoam vertex amplitude is the matrix element of $A$. (c) An example of LQG spinfoam amplitude made by connecting five quantum gates $A$. Post-selection and feed-forward make simulation of this spinfoam amplitude possible with a linear-optical quantum computer \cite{kok2007linear}.}
  \label{gate}
\end{figure}

The spinfoam amplitude is a network of quantum gates, which are quantum transitions of LQG quantum geometry states within Planck-scale volume regions \cite{han2019emergent}. Matrix elements of these quantum gates are called vertex amplitudes (see FIG.\ref{gate}). This feature of spinfoam amplitude shares a similarity with systems in quantum computation and allows spinfoams to be demonstrated on a quantum simulator device (see, e.g., \cite{li2019quantum,mielczarek2019spin} for existing studies relating LQG to quantum computation).

Here, we develop a new relation between the spinfoam LQG and a linear-optical quantum simulator. Based on this relation, we design a special-purpose linear-optical device for simulating spinfoam amplitudes (FIG. \ref{gate}). In our simulation, we map LQG quantum tetrahedron geometries to qubits and photon modes. We encode the spinfoam vertex amplitude in an optical quantum circuit, which is designed as a chain of linear-optical unitary operations followed by post-selection. This optical quantum circuit can be implemented on a chip, which is within the capability of current experiments \cite{wang2018multidimensional} and permits the simulation of spinfoam amplitudes with many vertices \---- due to the inherent scalability of linear-optical quantum information processors. Our work will shed light on computing spinfoam amplitudes with multiple vertices, which is intractable with classical numerical computation.


Our simulation has a broad applicability as it is valid for spinfoam amplitudes for all $j\geq 1/2$. It can also be applied to simulate tensor-network models that share similar structures as spinfoams and used to explore various aspects of topological quantum field theories.

\textit{Quantum tetrahedra and the Spinfoam vertex amplitude.}\----Among important quantum geometry states in LQG, a quantum tetrahedron is a tensor state $|\psi\rangle\in{\mathcal H}_{j_1}\otimes\cdots\otimes{\mathcal H}_{j_4}$ (${\mathcal H}_{j}$ is the SU(2)-irreducible representation labelled by spin-$j$). The state satisfies the following constraint equation
\begin{equation}
\left(\hat{\bf J}^{(1)}+\hat{\bf J}^{(2)}+\hat{\bf J}^{(3)}+\hat{\bf J}^{(4)}\right)|\psi\rangle=0,\label{closure}
\end{equation}
where $\hat{\bf J}^{(i)}=(\hat{J}_x,\hat{J}_y,\hat{J}_z)^{(i)}$ is the angular-momentum operator acting on ${\mathcal H}_{j_i}$. The $\hat{\bf J}^{(i)}$ quantizes the oriented area ${\bf E}^{(i)}=(E_x,E_y,E_z)^{(i)}$ of the $i^{\rm th}$ tetrahedron face ($i=1,\cdots,4$) \cite{barbieri1998quantum,rovelli2006semiclassical,li2019quantum} (see FIG.\ref{gate} (a)). $|{\bf E}^{(i)}|$ and ${\bf E}^{(i)}/|{\bf E}^{(i)}|$ are the area and unit normal of the $i^{\rm th}$ face. Eq.\ref{closure} quantizes the geometrical constraint ${\bf E}^{(1)}+{\bf E}^{(2)}+{\bf E}^{(3)}+{\bf E}^{(4)}=0$, meaning that the four tetrahedron faces form a closed surface. We denote by ${\mathcal H}_{\rm tet}$ the Hilbert space of all $|\psi\rangle$ satisfying Eq.\ref{closure}. ${\mathcal H}_{\rm tet}=\mathrm{Inv}_{SU(2)}(\mathcal{H}_{j_1}\otimes\cdots\otimes \mathcal{H}_{j_4})$ is the space of invariant tensors of SU(2). The spin $j_i$ is associated with the $i$-th face of the tetrahedron. Quantum tetrahedra are fundamental building blocks of quantum spatial geometries, since any geometry can be triangulated by tetrahedra. When $j_1=\cdots=j_4=1/2$, $\dim({\mathcal H}_{\rm tet})=2$, a quantum tetrahedron $|\psi\rangle$ can be described by a single qubit. 


Spinfoam amplitudes describe the evolution of quantum geometry states. The spinfoam amplitude is defined on a triangulation of a four-dimensional (4D) manifold, while its building block \---- vertex amplitude $A_\sigma$ \---- associates to a four-\emph{simplex} $\sigma$, the elementary cell of the 4D triangulation.

The four-simplex $\sigma$ is a 4D region whose boundary is a three-dimensional (3D) closed surface made by five tetrahedra (FIG.\ref{gate}(a)). We can choose to view $\sigma$ as a time evolution from three tetrahedra in the past to two tetrahedra in the future \footnote{Spinfoam is covariant so it can also be viewed in different perspectives e.g. from four tetrahedra in the past to one tetrahedron in the future.}. These tetrahedra carry quantum geometries $|\psi_i\rangle\in {\mathcal H}_{\rm tet}$ $(i=1,\cdots,5)$. $A_\sigma$ is a quantum transition amplitude from three initial quantum tetrahedra $|\psi_1\rangle,|\psi_2\rangle,|\psi_3\rangle$ to two final tetrahedra $|\psi_4\rangle,|\psi_5\rangle$. This quantum transition can be formulated as a quantum gate $A:{\mathcal H}_{\rm tet}\otimes{\mathcal H}_{\rm tet}\otimes{\mathcal H}_{\rm tet}\to{\mathcal H}_{\rm tet}\otimes{\mathcal H}_{\rm tet}$ (FIG.\ref{gate}(b)). The vertex amplitude $A_\sigma\equiv\langle\psi_4,\psi_5 |A|\psi_1,\psi_2,\psi_3\rangle$ is the probability amplitude of having an output $|\psi_4\rangle,|\psi_5\rangle$ provided the input is $|\psi_1\rangle,|\psi_2\rangle,|\psi_3\rangle$. A spinfoam amplitude in LQG is built by connecting $N$ quantum gates $A$, where each $A$ associates to a four-simplex $\sigma$ and $N$ is the number of $\sigma$'s in the triangulation (FIG.\ref{gate}(c)).

If we set all $j=1/2$, then $|\psi_1\rangle,\cdots,|\psi_5\rangle$ are qubits. Thus, $A$ is a quantum gate from three qubits to two qubits and can be simulated by a quantum linear-optical experiment. The design of the simulation is given in the next section. 

The simulation can be applied to higher spins ($j>1/2$) as well, with $\dim({\mathcal H}_{\rm tet})>2$. 
In concrete, let us firstly consider an example of $A$ whose quantum tetrahedra have all $j=1$, so that all five $\dim({\mathcal H}_{\rm tet})$ are 3-dimensional. We choose a basis $|e_{A=1,2,3}\rangle$ in each ${\mathcal H}_{\rm tet}$, and make the orthogonal decomposition 
\begin{eqnarray}
{\mathcal H}_{\rm tet}=\mathcal{H}^+\oplus\mathcal{H}^-,\quad \mathcal{H}^+\equiv {\mathcal H}^{(2D)}_{\rm tet}
\end{eqnarray}
where $\ch^+$ is spanned by $|e_1\rangle$ and $|e_2\rangle$, and $\ch^-$ is spanned by $|e_3\rangle$. We can restrict inputs and outputs of $A$ into subspaces $\ch^\pm$ and obtain sub-matrices. For instance, restricting all inputs and outputs of $A$ to five ${\mathcal H}^{+}$ gives $A^{+++++}$, whose matrix elements are $A^{+++++}_{AB,CDE}=\langle e_A,e_B|A|e_C,e_D,e_E\rangle$ where $A,B,C,D,E=1,2$. $A^{+++++}$ is a gate from 3 qubits to 2 qubits and is accordingly a $4\times 8$ matrix. Restricting some inputs and/or outputs to $\ch^-$ gives, e.g., $A^{+-+-+}$ whose matrix elements are $A^{+-+-+}_{A,CE}=\langle e_A,e_3|A|e_C,e_3,e_E\rangle$, where $A,C,E=1,2$. $A^{+-+-+}$ is a gate from 2 qubits to one qubit and is accordingly a $2\times 4$ matrix. All 32 $A^{a_1\cdots a_5}$ ($a_i=\pm$) are linear transformations of qubits ($A^{-----}$ a trivial transformation) and cover all information of $A$. Our strategy of linear-optical simulation is to design a quantum circuit on chip for each $A^{a_1\cdots a_5}$. We need 32 quantum circuits (less than 32 in practice since $A^{-----}$ is trivial and $A^{+----}$ is just a qubit) to simulate the complete $A$ with $j=1$. {The most nontrivial design for $A^{+++++}$ is discussed in detail in the next section, while all other circuits for $A^{\pm\pm\pm\pm\pm}$ are much simpler and can be designed similarly.} 

Our strategy can be easily generalized to $A$ with arbitrary $j$: $\ch_{\rm tet}$ of arbitrary dimension $d$ can be decomposed into mutually orthogonal subspaces $\ch^{(a)}$ with $\dim(\ch^{(a)})\leq2$:
\be
\ch_{tet}=\bigoplus_{a=1}^{M}\ch^{(a)},\quad M=\begin{cases}
\frac{d}{2}, & d\ \text{even}\\
\frac{d+1}{2}, & d\ \text{odd}
\end{cases}\label{hchoices}
\ee
Restricting inputs and outputs of $A$ in different $\ch^{(a)}$ gives $M^5$ quantum gates $A^{a_1\cdots a_5}$ ($a_i=1,\cdots,M$) of qubits. Each $A^{a_1\cdots a_5}$ can be cast into a linear-optical quantum circuit as discussed in the next section.

When we prepare the input state for $A$, we require the state to satisfy the area-matching condition \cite{perez2013spin}, i.e. for $\psi_m$ and $\psi_n$ corresponding to 2 tetrahedra sharing a face, their spin $j$ associated to the triangle has to be identical. This condition should also be imposed on the input state in general when we connect $N$ quantum gates $A$ to simulate spinfoams with $N$ 4-simplices.

In the following, we will discuss the implementation of the gate on a photonic chip. We note that such chips usually have room for active control, which may enable making all different configurations in one physical chip, without the need to fabricate a chip per gate.





\textit{Linear-optical simulator.}\----The gate $A$ from three qubits to two qubits can be represented by a $2^2\times 2^3$ matrix and is clearly non-unitary. However, it is possible to extend $A$ to a $12\times 12$ unitary matrix $U$ which includes $A$ as a submatrix \cite{vanmeter2007general}. We use the singular-value decomposition; $A=LSR$, where $L\,(4\times 4) ,\,R \,(8\times 8)$ are unitary and $S$ is $4\times 8$ matrix, with the singular values, $s_1,...,s_4$, on the diagonal and zeros elsewhere. 

\begin{figure}[t]
  \begin{center}
    \includegraphics[width = \linewidth]{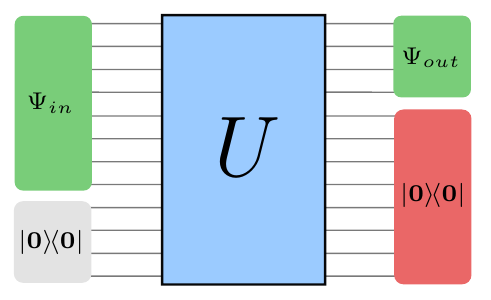}
  \end{center}
  \caption{The $12\times12$ unitary transformation. The input state is $1$ for one mode, $1,\cdots,8$ and $0$ for all other 11 modes. The output state is the evolved state in modes, $1,\cdots,4$ and conditioned vacuum is measured in modes, $5,\cdots,12$. }
  \label{unitary}
\end{figure}

Now we can reconstruct the unitary matrix:
\begin{equation}
U=
\begin{pmatrix}
A & L\sqrt{I-SS^\mathsf{T}}L \\
R\sqrt{I-S^\mathsf{T} S}R & -RS^\mathsf{T} L
\end{pmatrix}\,,
\label{Umat}
\end{equation}
where $I$ is the identity matrix (with the proper size). One can check that this $12\times 12$ matrix is unitary and that $A$ is submatrix of it. The condition for the unitarity of the matrix is that all of the singular values are strictly less than one \cite{fiedler2009suborthogonality}. This is indeed the case for spinfoam amplitudes. We obtain $A$ numerically for a Lorentzian Engle-Pereira-Rovelli-Livine (EPRL) spinfoam amplitude \cite{engle2008lqg,dona2019numerical}, and check it satisfies the condition (shown in the supplemental information \cite{supp1}). 
Generally speaking, all closed-system physical processes conserve probability. Thus, a complete physical process is unitary, and while partial physical process can be nonunitary, the singular values of such systems are limited by the complete process and cannot be larger than one \cite{fiedler2009suborthogonality}. We note that any unitary transformation can be implemented with linear optics, as explicitly shown in the supplemental information \cite{supp1}. 

\begin{figure*}[t]
  \begin{center}
    \includegraphics[width = 1\linewidth]{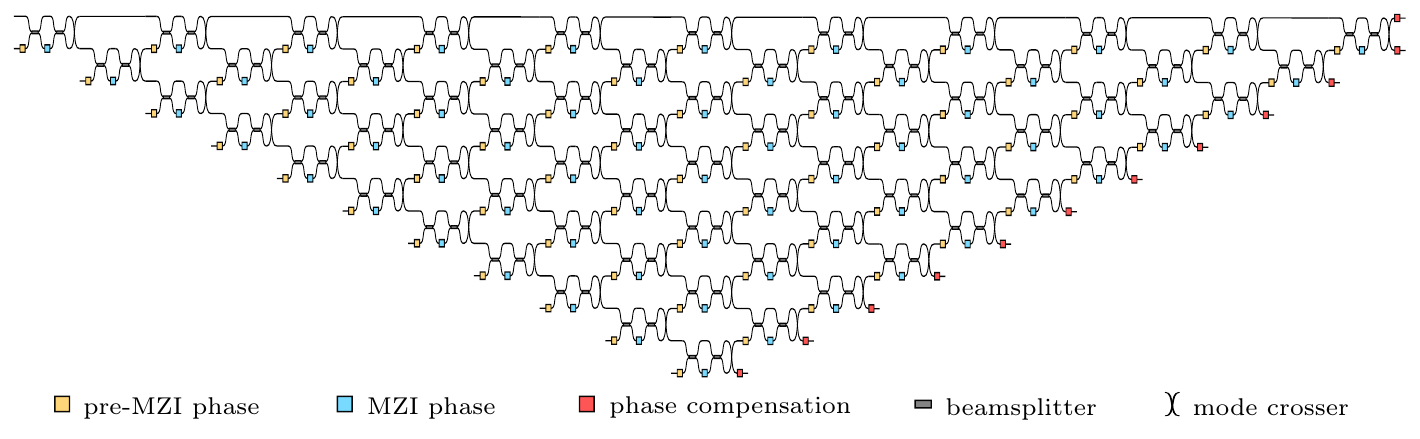}
  \end{center}
  \caption{Linear-optical circuit representation of $U$ (Eq. S5) for $N=12$ spatial modes. For consistency with notation, the top-left input corresponds to the $N^{\rm th}$ ($12^{\rm th}$) mode. The one below that is the $(N-1)^{\rm th}$ mode, and so on. The location of the corresponding output modes is found by following the transmission path of the input. For example, the bottom-right output is the $N^{\rm th}$ mode. The one above that is the $(N-1)^{\rm th}$ mode, and so on. The yellow boxes before the MZIs represent the phases $\phi$, the blue boxes within the MZIs represent the phases $\omega$, and the red boxes on the far right represent elements of the diagonal phase compensation matrix $D$. (see the supplemental information for more details \cite{supp1})}
  \label{fig-circuit}
\end{figure*}

This unitary can be implemented with any four-qubit system, where the $12 \times 12$ unitary above would be a submatrix of a $16 \times 16$ unitary acting on the four-qubit Hilbert space. We note that, in the general case of spins ($j> 1/2$), one can use four qudits with the same dimensionality as the spins. The decomposition of the unitary in this case, to two-qudit unitary operations, can be done in a similar fashion as the qubit case but is beyond the scope of this work.

We now show how a simpler decomposition can be done. Instead of using four qubits, one can use a single photon and 12 spatial modes. Not only is it much easier to conduct experiments with a single photon, but it is also easier to perform the two-mode operations between spatial-mode pairs. 
Let us rewrite the three input qubits as:

\begin{eqnarray}
&&\nonumber |\Psi\rangle_{\rm in} = |\psi\rangle_{1} \otimes |\psi\rangle_{2} \otimes |\psi\rangle_{3} =\\
&&\nonumber (\alpha_{10}|0\rangle + \alpha_{11}|1\rangle) \otimes (\alpha_{20}|0\rangle + \alpha_{21}|1\rangle) \otimes (\alpha_{20}|0\rangle + \alpha_{21}|1\rangle)\\
&&= \prod_{j=1}^3 \sum_{k=0}^1 \alpha_{jk}|k\rangle_j\,,
\label{SttIn1}
\end{eqnarray}
where $|k\rangle_j$ is the $k$ state of the $j$ qubit with amplitude of $\alpha_{jk}$. Rewriting the indices, the initial state is 

\begin{equation}
|\Psi\rangle_{\rm in} = \sum_{n=0}^7 \alpha_{n}|n\rangle\,,
\label{SttIn2}
\end{equation}
where $\alpha_{n}=\alpha_{1i}\alpha_{2j}\alpha_{3k}$ and $(ijk)$ is the binary representation of $n$. Thus $\Psi$ may be reinterpreted as a qudit of dimension $d=8$.

One implementation of this state is a single photon in a superposition of eight different spatial modes, i.e., eight different waveguides (Fig. \ref{unitary}). By taking such a system, the number of physical particles is reduced to one! The total unitary matrix can be implemented in a 12-waveguide chip where the number of integrated MZIs is bounded by $N(N-1)/2=66$ (Fig. \ref{fig-circuit}), which is within the capability of current optical experiments \cite{wang2018multidimensional}.

The elements of $A$ can be measured by changing the input state and monitoring the output. In general, the output state is:

\begin{equation}
|\bar \Psi\rangle_{\rm out}= \hat U|\bar \Psi\rangle_{\rm in}\,,
\label{evlv}
\end{equation}
where the $\bar \cdot$ denotes the complete 12-mode states in contrast to the 8- or 4-mode reduced states. Taking the initial state to be the $j^{\rm th}$ basis vector of the trivial basis, which physically means to input one photon in the $j^{\rm th}$ port, will result the output state amplitudes to be the $j^{\rm th}$ column of U:

\begin{equation}
U
\begin{pmatrix}
0\\
\vdots\\
0\\
1\\
\vdots\\
0
\end{pmatrix}
\begin{matrix}
\\
\\
\\
j^{\rm th}\\
\\
\\
\end{matrix}
=
\begin{pmatrix}
U_{1j}\\
U_{2j}\\
\\
\vdots\\
\\
U_{Nj}
\end{pmatrix}\,,
\label{evlv2}
\end{equation}
Measuring the detection probability of the photon at the $i^{\rm th}$ output port then gives the elements of $U$:

\begin{equation}
P_{ij} = |\langle i | \hat U | j \rangle |^2 = |U_{ij}|^2 \,.
\label{Umat2}
\end{equation}
Recalling Eq. \ref{Umat}, $A$ is a submatrix of $U$, thus Eq. \ref{Umat2} holds also for $A$. 
The phase of $A$ can be found by preparing the initial state in an equal superposition of two modes, say $j$ and $j'$. Then, the detection probability of the photon in the $i^{\rm th}$ mode is $|U_{ij}+U_{ij'}|^2/2$. Taking this and Eq. \ref{Umat2}, the phase between $U_{ij}$ and $U_{ij'}$ can be extracted.     

The protocol for implementing $A$ includes post-selection of vacuum in modes $5,\cdots,12$ (see e.g. Fig. \ref{unitary} and Ref. \cite{vanmeter2007general}). Here we see another advantage of using just one physical particle; if it is measured in one mode it cannot be measured in any other modes. Therefore, the post-selection is automatically satisfied by just measuring the detection probability of the four first modes and ignoring any photon in the other modes.

Unlike other implementations to LQG, ours also includes path entanglement, which is generated by the beamsplitters and  post selection \cite{vanmeter2007general}. We quantify the entanglement with the Von Neumann entropy: $-\sum_{n} \tilde s_n \log_2 \tilde s_n$, where $\tilde s_n$ is the is the $n^{\rm th}$ singular value of the density matrix after post selection and partial trace. Since there are four spatial modes, there are 14 different ways to perform the partial trace, and we maximized the entropy over all options.  
The amount of entanglement depends on the gate A and the input state. Taking an example gate (see supplemental information \cite{supp1}) and varying in the input state, the entanglement ranges between $.014-.986$, with $1$ being maximal for the four-dimensional Hilbert space (see supplemental information for more details \cite{supp1}). 



The above discussion is for encoding the spinfoam vertex amplitude in one chip of optical gates. The generalization to spinfoam amplitudes with $N$ vertex amplitudes is made by building $N$ similar optical chips and connecting them optically. Performing measurement on this enlarged system will produce spinfoam amplitudes with $N$ vertices. Implementing the gate-on-chip with 12 spatial modes reduces the required number of photons from four, without spatial multiplexing, to one with it. Thus, the number of photons is bound by the number of vertices and thus simulating a few-vertex spinfoam amplitude is experimentally practical. 

Given a spinfoam amplitude with $N$ vertices and spins $\{j_f\}$, its complexity can be estimated: We denote by $\mathcal{C}$ the complexity of a single $U$. $\mathcal{C}$ is the number of 2-mode gates in $U$ \cite{barenco1995elementary}, and is bounded by $66$ (see the supplemental information \cite{supp1}). $N$ spinfoam vertices give the complexity $\mathcal{C}^N$. Moreover there are multiple choices of $\ch^{(a)}$ for $j>1/2$ at each tetrahedron $\Delta$ as described in (\ref{hchoices}). If the number of choices at each $\Delta$ is denoted by $M_\Delta$, the total complexity of a spinfoam amplitude is bounded by $\mathcal{C}^N\prod_{\Delta}M_\Delta$ where $\prod_{\Delta}$ products over all tetrahedra $\Delta$ in the 4D triangulation. If we take into account summing over internal spins in the spinfoam amplitude, the complexity is bounded by $\mathcal{C}^N\prod_{\Delta}M_\Delta \prod_fJ_f$ where $\prod_f$ products over all internal triangles $f$ and $J_f$ is the number of spins summed at $f$. In principle, the sum over all triangulations should be calculated, but summing over triangulations is beyond the scope of the present letter.

\textit{Summary.}\----In summary, we have developed a scalable linear-optical implementation for efficiently simulating LQG spinfoam vertex amplitudes -- a problem which is strongly believed to be in the computational complexity class BQP -- which means there exists no efficient classical simulation. The implementation of the quantum gate that simulates the vertex amplitude requires only a single photon and a 12 spatial-mode circuit. The extension to N-vertex spinfoam amplitudes can then be made by `stitching' many of these primitive vertex-amplitude gates together (Fig. \ref{gate}). Thus, simulating N-vertex spinfoam amplitudes in LQG is now within experimental reach.  


\textit{Acknowledgments.}\----MH and DQ receives support from the National Science Foundation through grant PHY-1912278. AJB, LC, and JPD acknowledge support from the Army Research Office, the Air Force Office of Scientific Research, and the National Science Foundation.   

\nocite{clements2016optimal}
\bibliography{linear_optical_LQG}

\end{document}